\documentclass[]{interact}

\usepackage{epstopdf}
\usepackage{subfigure}

\usepackage{natbib}
\bibpunct[, ]{(}{)}{;}{a}{}{,}

\theoremstyle{plain}

\theoremstyle{definition}

\theoremstyle{remark}

\usepackage{lscape}
\usepackage{hyperref}
\usepackage[utf8]{inputenc}

\usepackage{setspace}
\usepackage{booktabs}
\usepackage{longtable}
\usepackage{array}
\usepackage{multirow}
\usepackage{wrapfig}
\usepackage{float}
\usepackage{colortbl}
\usepackage{pdflscape}
\usepackage{tabu}
\usepackage{threeparttable}
\usepackage{threeparttablex}
\usepackage[normalem]{ulem}
\usepackage{makecell}
\usepackage{xcolor}

\begin{document}

\articletype{}

\title{A Plot is Worth a Thousand Tests: Assessing Residual Diagnostics
with the Lineup Protocol}

\author{\name{Weihao Li$^{a}$, Dianne Cook$^{a}$, Emi Tanaka$^{a, b,
c}$, Susan VanderPlas$^{d}$}
\affil{$^{a}$Department of Econometrics and Business Statistics, Monash
University, Clayton, VIC, Australia; $^{b}$Biological Data Science
Institute, Australian National University, Acton, ACT,
Australia; $^{c}$Research School of Finance, Actuarial Studies and
Statistics, Australian National University, Acton, ACT,
Australia; $^{d}$Department of Statistics, University of Nebraska,
Lincoln, Nebraska, USA}
}

\thanks{CONTACT Weihao
Li. Email: \href{mailto:weihao.li@monash.edu}{\nolinkurl{weihao.li@monash.edu}}, Dianne
Cook. Email: \href{mailto:dicook@monash.edu}{\nolinkurl{dicook@monash.edu}}, Emi
Tanaka. Email: \href{mailto:emi.tanaka@anu.edu.au}{\nolinkurl{emi.tanaka@anu.edu.au}}, Susan
VanderPlas. Email: \href{mailto:susan.vanderplas@unl.edu}{\nolinkurl{susan.vanderplas@unl.edu}}}

\maketitle

\begin{abstract}
Regression experts consistently recommend plotting residuals for model
diagnosis, despite the availability of many numerical hypothesis test
procedures designed to use residuals to assess problems with a model
fit. Here we provide evidence for why this is good advice using data
from a visual inference experiment. We show how conventional tests are
too sensitive, which means that too often the conclusion would be that
the model fit is inadequate. The experiment uses the lineup protocol
which puts a residual plot in the context of null plots. This helps
generate reliable and consistent reading of residual plots for better
model diagnosis. It can also help in an obverse situation where a
conventional test would fail to detect a problem with a model due to
contaminated data. The lineup protocol also detects a range of
departures from good residuals simultaneously. Supplemental materials
for the article are available online.
\end{abstract}


\section{Introduction}\label{introduction}

\begin{quote}
\emph{``Since all models are wrong the scientist must be alert to what
is importantly wrong.''} \citep{box1976science}
\end{quote}

Diagnosing a model is an important part of building an appropriate
model. In linear regression analysis, studying the residuals from a
model fit is a common diagnostic activity. Residuals summarise what is
not captured by the model, and thus provide the capacity to identify
what might be wrong.

We can assess residuals in multiple ways. To examine the univariate
distribution, residuals may be plotted as a histogram or normal
probability plot. Using the classical normal linear regression model as
an example, if the distribution is symmetric and unimodal, we would
consider it to be well-behaved. However, if the distribution is skewed,
bimodal, multimodal, or contains outliers, there would be cause for
concern. We can also inspect the distribution by conducting a
goodness-of-fit test, such as the Shapiro-Wilk normality test
\citep{shapiro1965analysis}.

Scatterplots of residuals against the fitted values, and each of the
explanatory variables, are commonly used to scrutinize their
relationships. If there are any visually discoverable associations, the
model is potentially inadequate or incorrectly specified. We can also
potentially discover patterns not directly connected to a linear model
assumption from these residual plots, such as the discreteness or
skewness of the fitted values, and outliers. To read residual plots, one
looks for noticeable departures from the model such as non-linear
pattern or heteroskedasticity. A non-linear pattern would suggest that
the model needs to have some additional non-linear terms.
Heteroskedasticity suggests that the error is dependent on the
predictors, and hence violates the independence assumption. Statistical
tests were developed to provide objective assessment, for example, of
non-linear patterns \citep[e.g.][]{ramsey1969tests}, and
heteroskedasticity \citep[e.g.][]{breusch1979simple}.

The common wisdom of experts is that plotting the residuals is
indispensable for diagnosing model fits
\citep{draper1998applied, cook1982residuals, montgomery1982introduction}.
The lack of empirical evidence for the ubiquitous advice is
\emph{curious}, and is what this article tackles.

Additionally, relying solely on the subjective assessment of a single
plot can be problematic. People will almost always see a pattern
\citep[see][]{kahneman2011thinking}, so the question that really needs
answering is whether any pattern perceived is consistent with
randomness, or sampling variability, or noise. Correctly judging whether
\emph{no} pattern exists in a residual plot is a difficult task.
\cite{loy2021bringing} emphasizes that this is especially difficult to
teach to new analysts and students, and advocates to the broader use of
the lineup protocol \citep{bujastatistical2009}.

\begin{figure}[t!]

{\centering \includegraphics[width=1\linewidth]{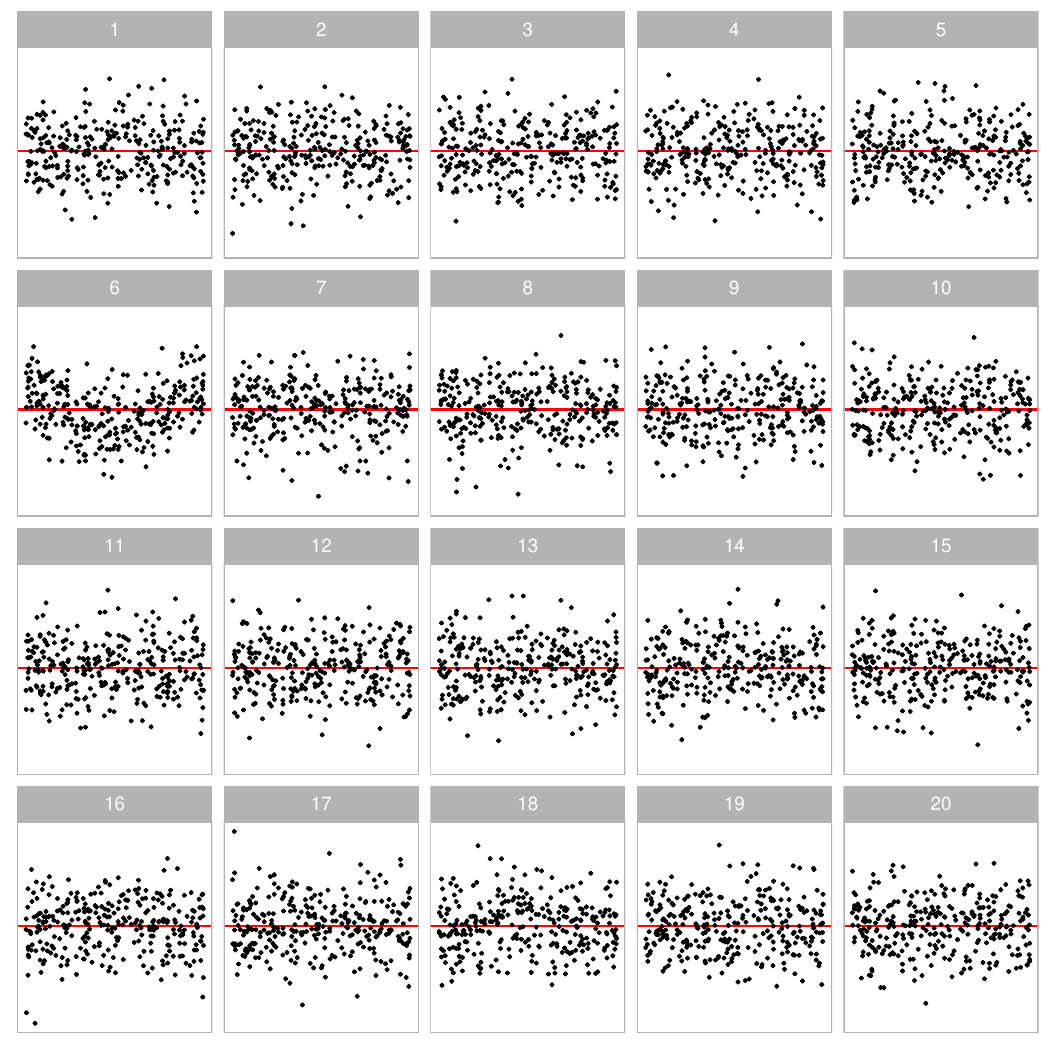} 

}

\caption{Visual testing is conducted using a lineup, as in the example here. The residual plot computed from the observed data is embedded among 19 null plots, where the residuals are simulated from a standard error model. Computing the $p$-value requires that the lineup be examined by a number of human judges, each asked to select the most different plot. A small $p$-value would result from a substantial number selecting the data plot (at position $6$, exhibiting non-linearity).}\label{fig:first-example-lineup}
\end{figure}

The lineup protocol places a data plot in a field of null plots,
allowing for a comparison of patterns due purely by chance to what is
perceived in the data plot. For residual analysis this is especially
helpful for gauging whether there is \emph{no} pattern. (Figure
\ref{fig:first-example-lineup} shows an example of a lineup of residual
plots.) In its strict use, one would insist that the data plot is not
seen before seeing the lineup, so that the observer does not know which
is the true plot. When used this way, it provides an objective test for
data plots. \cite{majumdervalidation2013} validated that results from
lineups assessed by human observers performed similarly to conventional
tests. One would not use a lineup when a conventional test exists and is
adequate because it is more manually expensive to conduct. However,
where no adequate conventional test exists, it is invaluable, as shown
by \cite{loy2013diagnostic}. Here we use the lineup as a vehicle to
rigorously explore why experts advise that residual plots are
indispensable despite the prevalence of numerical tests.

The paper is structured as follows. Section \ref{background} describes
the background on the types of departures that one expects to detect,
and outlines a formal statistical process for reading residual plots,
called visual inference. Section \ref{significance-calculation}
describes the calculation of the statistical significance and power of
the test. Section \ref{experimental-design} details the experimental
design to compare the decisions made by formal hypothesis testing, and
how humans would read diagnostic plots. The results are reported in
Section \ref{results}. We conclude with a discussion of the presented
work, and ideas for future directions.

\section{Background}\label{background}

\subsection{Departures from good residual
plots}\label{departures-from-good-residual-plots}

\begin{figure}[t!]

{\centering \includegraphics[width=1\linewidth]{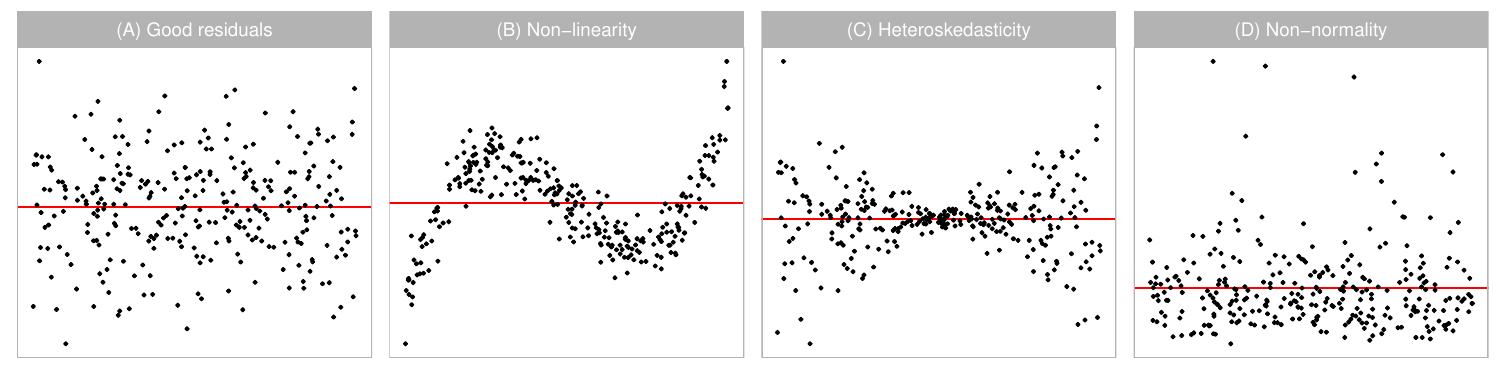} 

}

\caption{Example residual vs fitted value plots (horizontal line indicates 0): (A) classically good looking residuals, (B) non-linear pattern indicates that the model has not captured a non-linear association, (C) heteroskedasticity indicating that variance around the fitted model is not uniform, and (D) non-normality where the residual distribution is not symmetric around 0. The latter pattern might best be assessed using a univariate plot of the residuals, but patterns B and C need to be assessed using a residual vs fitted value plot.}\label{fig:residual-plot-common-departures}
\end{figure}

Graphical summaries where residuals are plotted against fitted values,
or other functions of the predictors (expected to be approximately
orthogonal to the residuals) are considered to be the most important
residual plots by \citet{cook1999applied}. Figure
\ref{fig:residual-plot-common-departures}A shows an example of an ideal
residual plot where points are symmetrically distributed around the
horizontal zero line (red), with no discernible patterns. There can be
various types of departures from this ideal pattern. Non-linearity,
heteroskedasticity and non-normality, shown in Figures
\ref{fig:residual-plot-common-departures}B,
\ref{fig:residual-plot-common-departures}C, and
\ref{fig:residual-plot-common-departures}D, respectively, are three
commonly checked departures.

Model misspecification occurs if functions of predictors that needed to
accurately describe the relationship with the response are incorrectly
specified. This includes instances where a higher-order polynomial term
of a predictor is wrongfully omitted. Any non-linear pattern visible in
the residual plot could be indicative of this problem. An example
residual plot containing visual pattern of non-linearity is shown in
Figure \ref{fig:residual-plot-common-departures}B. One can clearly
observe the ``S-shape'' from the residual plot, which corresponds to the
cubic term that should have been included in the model.

Heteroskedasticity refers to the presence of non-constant error variance
in a regression model. It indicates that the distribution of residuals
depends on the predictors, violating the independence assumption. This
can be seen in a residual plot as an inconsistent spread of the
residuals relative to the fitted values or predictors. An example is the
``butterfly'' shape shown in Figure
\ref{fig:residual-plot-common-departures}C, or a ``left-triangle'' and
``right-triangle'' shape where the smallest variance occurs at one side
of the horizontal axis.

Figure \ref{fig:residual-plot-common-departures}D shows a scatterplot
where the residuals have a skewed distribution, as seen by the uneven
vertical spread. Unlike non-linearity and heteroskedasticity,
non-normality is usually detected with a different type of residual
plot: a histogram or a normal probability plot. Because we focus on
scatterplots, non-normality is not one of the departures examined in
depth in this paper. \citep[ discuss related work on non-normality
checking.]{loy2016variations}

\subsection{Conventionally testing for
departures}\label{conventionally-testing-for-departures}

Many different hypothesis tests are available to detect specific model
defects. For example, the presence of heteroskedasticity can usually be
tested by applying the White test \citep{white1980heteroskedasticity} or
the Breusch-Pagan (BP) test \citep{breusch1979simple}, which are both
derived from the Lagrange multiplier test \citep{silvey1959lagrangian}
principle that relies on the asymptotic properties of the null
distribution. To test specific forms of non-linearity, one may apply the
F-test as a model structural test to examine the significance of a
specific polynomial and non-linear forms of the predictors, or the
significance of proxy variables as in the Ramsey Regression Equation
Specification Error Test (RESET) \citep{ramsey1969tests}. The
Shapiro-Wilk (SW) normality test \citep{shapiro1965analysis} is the most
widely used test of non-normality included by many of the statistical
software programs. The Jarque-Bera test \citep{jarque1980efficient} is
also used to directly check whether the sample skewness and kurtosis
match a normal distribution.

Table \ref{tab:example-residual-plot-table} displays the \(p\)-values
from the RESET, BP and SW tests applied to the residual plots in Figure
\ref{fig:residual-plot-common-departures}. The RESET test and BP test
were computed using the \texttt{resettest} and \texttt{bptest} functions
from the R package \texttt{lmtest}, respectively. The SW test was
computed using the \texttt{shapiro.test} from the core R package
\texttt{stats}. \footnote{Although we did not use it, it is useful to
  know that the R package \texttt{skedastic} \citep{skedastic} also
  contains a large collection of functions to test for
  heteroskedasticity.} The RESET test requires the selection of a power
parameter. \citet{ramsey1969tests} recommends a power of four, which we
adopted in our analysis.

For residual plots in Figure \ref{fig:residual-plot-common-departures},
we would expect the RESET test for non-linearity to reject residual plot
B, the BP test for heteroskedasticity to reject the residual plot C, and
the SW test for non-normality to reject residual plot D, which they all
do and all tests also correctly fail to reject residual plot A.
Interestingly, the BP and SW tests also reject the residual plots
exhibiting structure that they were not designed for.
\citet{cook1982residuals} explain that most residual-based tests for a
particular departure from the model assumptions are also sensitive to
other types of departures. This could be considered a Type III error
\citep{kimball1957errors}, where the null hypothesis of good residuals
is correctly rejected but for the wrong reason. Also, some types of
departure can have elements of other types of departure, for example,
non-linearity can appear like heteroskedasticity. Additionally, other
data problems such as outliers can trigger rejection (or not) of the
null hypothesis \citep{cook1999applied}.

With large sample sizes, hypothesis tests may reject the null hypothesis
when there is only a small effect. (A good discussion can be found in
\citet{kirk1996practical}.) While such rejections may be statistically
correct, their sensitivity may render the results impractical. A key
goal of residual plot diagnostics is to identify potential issues that
could lead to incorrect conclusions or errors in subsequent analyses,
but minor defects in the model are unlikely to have a significant impact
and may be best disregarded for practical purposes. The experiment
discussed in this paper specifically addresses this tension between
statistical significance and practical significance.

\begin{table}

\caption{\label{tab:example-residual-plot-table}Statistical significance testing for departures from good residuals for plots in Figure \ref{fig:residual-plot-common-departures}. Shown are the $p$-values calculated for the RESET, the BP and the SW tests. The good residual plot (A) is judged a good residual plot, as expected, by all tests. The non-linearity (B) is detected by all tests, as might be expected given the extreme structure.}
\centering
\begin{tabular}[t]{llrrr}
\toprule
Plot & Departures & RESET & BP & SW\\
\midrule
A & None & 0.779 & 0.133 & 0.728\\
B & Non-linearity & \em{0.000} & \em{0.000} & \em{0.039}\\
C & Heteroskedasticity & 0.658 & \em{0.000} & \em{0.000}\\
D & Non-normality & 0.863 & 0.736 & \em{0.000}\\
\bottomrule
\end{tabular}
\end{table}

\subsection{Visual test procedure based on
lineups}\label{visual-test-procedure-based-on-lineups}

The examination of data plots to infer signals or patterns (or lack
thereof) is fraught with variation in the human ability to interpret and
decode the information embedded in a graph
\citep{cleveland1984graphical}. In practice, over-interpretation of a
single plot is common. For instance, \citet{roy2015using} described a
published example where authors over-interpreted separation between gene
groups from a two-dimensional projection of a linear discriminant
analysis even when there were no differences in the expression levels
between the gene groups.

One solution to over-interpretation is to examine the plot in the
context of natural sampling variability assumed by the model, called the
lineup protocol, as proposed in \citet{buja2009statistical}.
\citet{majumder2013validation} showed that the lineup protocol is
analogous to the null hypothesis significance testing framework. The
protocol consists of \(m\) randomly placed plots, where one plot is the
data plot, and the remaining \(m - 1\) plots, referred to as the
\emph{null plots}, are constructed using the same graphical procedure as
the data plot but the data is replaced with null data that is generated
in a manner consistent with the null hypothesis, \(H_0\). Then, an
observer who has not seen the data plot is asked to point out the most
different plot from the lineup. Under \(H_0\), it is expected that the
data plot would have no distinguishable difference from the null plots,
and the probability that the observer correctly picks the data plot is
\(1/m\). If one rejects \(H_0\) as the observer correctly picks the data
plot, then the Type I error of this test is \(1/m\). This protocol
requires a priori specification of \(H_0\) (or at least a null data
generating mechanism), much like the requirement of knowing the sampling
distribution of the test statistic in null hypothesis significance
testing framework.

Figure \ref{fig:first-example-lineup} is an example of a lineup
protocol. If the data plot at position \(6\) is identifiable, then it is
evidence for the rejection of \(H_0\). In fact, the actual residual plot
is obtained from a misspecified regression model with missing non-linear
terms.

Data used in the \(m - 1\) null plots needs to be simulated. In
regression diagnostics, sampling data consistent with \(H_0\) is
equivalent to sampling data from the assumed model. As
\citet{buja2009statistical} suggested, \(H_0\) is usually a composite
hypothesis controlled by nuisance parameters. Since regression models
can have various forms, there is no general solution to this problem,
but it sometimes can be reduced to a so called ``reference
distribution'' by applying one of the three methods: (i) sampling from a
conditional distribution given a minimal sufficient statistic under
\(H_0\), (ii) parametric bootstrap sampling with nuisance parameters
estimated under \(H_0\), and (iii) Bayesian posterior predictive
sampling. The conditional distribution given a minimal sufficient
statistic is the best justified reference distribution among the three
\citep{buja2009statistical}. Under this method, the null residuals can
essentially be simulated by independent drawing from a standard normal
random distribution, then regressing the draws on the predictors, and
then re-scaling it by the ratio of the residual sum of square in two
regressions.

The effectiveness of lineup protocol for regression analysis has been
validated by \citet{majumder2013validation} under relatively simple
settings with up to two predictors. Their results suggest that visual
tests are capable of testing the significance of a single predictor with
a similar power to a t-test, though they express that in general it is
unnecessary to use visual inference if there exists a corresponding
conventional test, and they do not expect the visual test to perform
equally well as the conventional test. In their third experiment, where
the contamination of the data violate the assumptions of the
conventional test, visual test outperforms the conventional test by a
large margin. This supports the use of visual inference in situations
where there are no existing numerical testing procedures. Visual
inference has also been integrated into diagnostics for hierarchical
linear models where the lineup protocol is used to judge the assumptions
of linearity, normality and constant error variance for both the level-1
and level-2 residuals
\citep{loy2013diagnostic, loy2014hlmdiag, loy2015you}.

\section{Calculation of statistical significance and test
power}\label{significance-calculation}

\subsection{What is being tested?}\label{what-is-being-tested}

In diagnosing a model fit using the residuals, we are generally
interested in testing whether ``\emph{the regression model is correctly
specified}'' (\(H_0\)) against the broad alternative ``\emph{the
regression model is misspecified}'' (\(H_a\)). However, it is
practically impossible to test this broad \(H_0\) with conventional
tests, because they need specific structure causing the departure to be
quantifiable in order to be computable. For example, the RESET test for
detecting non-linear departures is formulated by fitting
\(y = \tau_0 + \sum_{i=1}^{p}\tau_px_p +\gamma_1\hat{y}^2 + \gamma_2\hat{y}^3 + \gamma_3\hat{y}^4 + u, ~~u \sim N(0, \sigma_u^2)\)
in order to test \(H_0:\gamma_1 = \gamma_2 = \gamma_3 = 0\) against
\(H_a: \gamma_1 \neq 0 \text{ or } \gamma_2 \neq 0 \text{ or } \gamma_3 \neq 0\).
Similarly, the BP test is designed to specifically test \(H_0:\)
\emph{error variances are all equal}
(\(\zeta_i=0 \text{ for } i=1,..,p\)) versus the alternative \(H_a:\)
\emph{that the error variances are a multiplicative function of one or
more variables} (\(\text{at least one } \zeta_i\neq 0\)) from
\(e^2 = \zeta_0 + \sum_{i=1}^{p}\zeta_i x_i + u, ~ u\sim N(0,\sigma_u^2)\).

While a battery of conventional tests for different types of departures
could be applied, this is intrinsic to the lineup protocol. The lineup
protocol operates as an omnibus test, able to detect a range of
departures from good residuals in a single application.

\subsection{\texorpdfstring{Statistical
significance\label{sig}}{Statistical significance}}\label{statistical-significance}

In hypothesis testing, a \(p\)-value is defined as the probability of
observing test results at least as extreme as the observed result
assuming \(H_0\) is true. Conventional hypothesis tests usually have an
existing method to derive or compute the \(p\)-value based on the null
distribution. The method to estimate a \(p\)-value for a visual test
essentially follows the process detailed by
\citet{vanderplas2021statistical}. Details are given in Appendix A.

\subsection{Power of the tests}\label{power-of-the-tests}

The power of a model misspecification test is the probability that
\(H_0\) is rejected given the regression model is misspecified in a
specific way. It is an important indicator when one is concerned about
whether model assumptions have been violated. In practice, one might be
more interested in knowing how much the residuals deviate from the model
assumptions, and whether this deviation is of practical significance.

The power of a conventional hypothesis test is affected by both the true
parameters \(\boldsymbol{\theta}\) and the sample size \(n\). These two
can be quantified in terms of effect size \(E\) to measure the strength
of the residual departures from the model assumptions. Details about the
calculation of effect size are provided in Section \ref{effect-size}
after the introduction of the simulation model used in our experiment.
The theoretical power of a test is sometimes not a trivial solution, but
it can be estimated if the data generating process is known. We use a
predefined model to generate a large set of simulated data under
different effect sizes, and record if the conventional test rejects
\(H_0\). The probability of the conventional test rejects \(H_0\) is
then fitted by a logistic regression formulated as

\vspace{-\baselineskip}

\begin{equation} \label{eq:logistic-regression-1-1}
Pr(\text{reject}~H_0|H_1,E) = \Lambda\left(\log\left(\frac{0.05}{0.95}\right) + \beta_1 E\right),
\end{equation}

\noindent where \(\Lambda(.)\) is the standard logistic function given
as \(\Lambda(z) = \exp(z)(1+\exp(z))^{-1}\). The effect size \(E\) is
the only predictor and the intercept is fixed to \(\log(0.05/0.95)\) so
that \(\hat{Pr}(\text{reject}~H_0|H_1,E = 0) = 0.05\), the desired
significance level.

The power of a visual test on the other hand, may additionally depend on
the ability of the particular participant, as the skill of each
individual may affect the number of observers who identify the data plot
from the lineup \citep{majumder2013validation}. To address this issue,
\citet{majumder2013validation} models the probability of participant
\(j\) correctly picking the data plot from lineup \(l\) using a
mixed-effect logistic regression, with participants treated as random
effects. Then, the estimated power of a visual test evaluated by a
single participant is the predicted value obtained from the mixed
effects model. However, this mixed effects model does not work with
scenario where participants are asked to select one or more most
different plots. In this scenario, having the probability of a
participant \(j\) correctly picking the data plot from a lineup \(l\) is
insufficient to determine the power of a visual test because it does not
provide information about the number of selections made by the
participant for the calculation of the \(p\)-value. Therefore, we
directly estimate the probability of a lineup being rejected by assuming
that individual skill has negligible effect on the variation of the
power. This assumption essentially averages out the subject ability and
helps to simplify the model structure, thereby obviating a costly
large-scale experiment to estimate complex covariance matrices. The same
model given in Equation \ref{eq:logistic-regression-1-1} is applied to
model the power of a visual test.

To study various factors contributing to the power of both tests, the
same logistic regression model is fit on different subsets of the
collated data grouped by levels of factors. These include the
distribution of the fitted values, type of the simulation model and the
shape of the residual departures.

\section{Experimental design}\label{experimental-design}

Our experiment was conducted over three data collection periods to
investigate the difference between conventional hypothesis testing and
visual inference in the application of linear regression diagnostics.
Two types of departures, non-linearity and heteroskedasticity, were
collected during data collection periods I and II. The data collection
period III was designed primarily to measure human responses to null
lineups so that the visual \(p\)-values can be estimated. Additional
lineups for both non-linearity and heteroskedasticity, using uniform
fitted value distributions, were included for additional data, and to
avoid participant frustration of too many difficult tasks.

During the experiment, every participant recruited from the Prolific
crowd-sourcing platform \citep{palan2018prolific} was presented with a
block of 20 lineups. A lineup consisted of a randomly placed data plot
and 19 null plots, which were all residual plots drawn with raw
residuals on the y-axis and fitted values on the x-axis. An additional
horizontal red line was added at \(y = 0\) as a visual reference. The
data in the data plot was simulated from one of two models described in
Section \ref{simulating-departures-from-good-residuals}, while the data
of the remaining 19 null plots were generated by the residual rotation
technique discussed in Section
\ref{visual-test-procedure-based-on-lineups}.

In each lineup evaluation, the participant was asked to select one or
more plots that are most different from others, provide a reason for
their selections, and evaluate how different they think the selected
plots are from others. If there is no noticeable difference between
plots in a lineup, participants had the option to select zero plots
without the need to provide a reason. During the process of recording
the responses, a zero selection was considered to be equivalent to
selecting all 20 plots. No participant was shown the same lineup twice.
Information about preferred pronouns, age group, education, and previous
experience in visual experiments were also collected. A participant's
submission was only included in the analysis if the data plot is
identified for at least one attention check.

Overall, we collected 7974 evaluations on 1152 unique lineups performed
by 443 participants. A summary of the factors used in the experiment can
be found in Table \ref{tab:model-factor-table}. There were four levels
of the non-linear structure, and three levels of heteroskedastic
structure. The signal strength was controlled by error variance
(\(\sigma\)) for the non-linear pattern, and by a ratio (\(b\))
parameter for the heteroskedasticity. Additionally, three levels of
sample size (\(n\)) and four different fitted value distributions were
incorporated.

\begin{table}

\caption{\label{tab:model-factor-table}Levels of the factors used in data collection periods I, II, and III.}
\centering
\resizebox{\linewidth}{!}{
\begin{tabular}[t]{rr|rr|rcrr|rr|rcrr|rr|rcrr|rr|rcrr|rr|rcrr|rr|rc}
\toprule
\multicolumn{2}{c}{Non-linearity} & \multicolumn{2}{c}{Heteroskedasticity} & \multicolumn{2}{c}{Common} \\
\cmidrule(l{3pt}r{3pt}){1-2} \cmidrule(l{3pt}r{3pt}){3-4} \cmidrule(l{3pt}r{3pt}){5-6}
\multicolumn{1}{c}{Poly Order ($j$)} & \multicolumn{1}{c}{SD ($\sigma$)} & \multicolumn{1}{c}{Shape ($a$)} & \multicolumn{1}{c}{Ratio ($b$)} & \multicolumn{1}{c}{Size ($n$)} & \multicolumn{1}{c}{Distribution of the fitted values} \\
\cmidrule(l{3pt}r{3pt}){1-1} \cmidrule(l{3pt}r{3pt}){2-2} \cmidrule(l{3pt}r{3pt}){3-3} \cmidrule(l{3pt}r{3pt}){4-4} \cmidrule(l{3pt}r{3pt}){5-5} \cmidrule(l{3pt}r{3pt}){6-6}
2 & 0.25 & -1 & 0.25 & 50 & Uniform\\
3 & 1.00 & 0 & 1.00 & 100 & Normal\\
6 & 2.00 & 1 & 4.00 & 300 & Skewed\\
18 & 4.00 &  & 16.00 &  & Discrete\\
 &  &  & 64.00 &  & \\
\bottomrule
\end{tabular}}
\end{table}

\subsection{Simulating departures from good
residuals}\label{simulating-departures-from-good-residuals}

\subsubsection{Non-linearity and
Heteroskedasticity}\label{non-linearity-and-heteroskedasticity}

\begin{figure}[!h]

{\centering \includegraphics[width=1\linewidth]{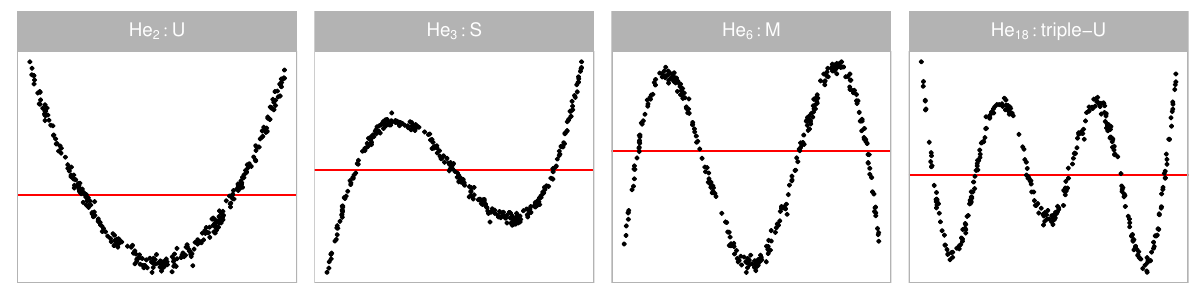} 

}

\caption{Polynomial forms generated for the residual plots used to assess detecting non-linearity. The four shapes are generated by varying the order of polynomial given by $j$ in $He_j(.)$.}\label{fig:different-shape-of-herimite}
\end{figure}

\begin{figure}[!h]

{\centering \includegraphics[width=1\linewidth]{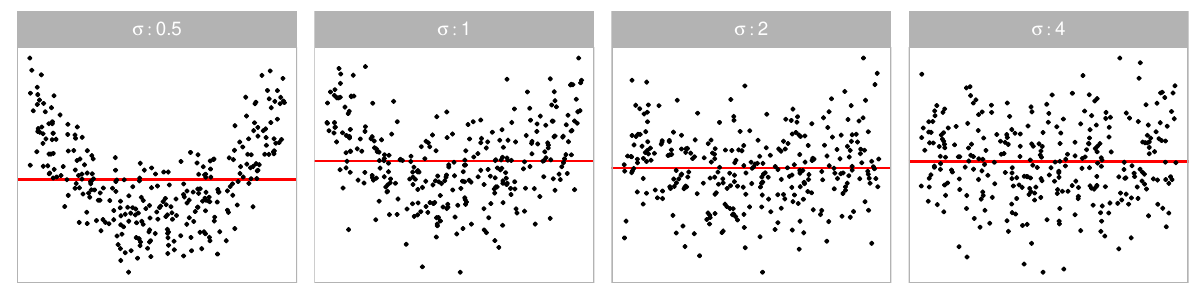} 

}

\caption{Examining the effect of $\sigma$ on the signal strength in the non-linearity detection, for $n=300$, uniform fitted value distribution and the ``U" shape. As $\sigma$ increases the signal strength decreases, to the point that the ``U" is almost unrecognisable when $\sigma=4$.}\label{fig:different-sigma}
\end{figure}

Data collection period I was designed to study the ability of
participants to detect non-linearity departures from residual plots. The
non-linearity departure was constructed by omitting a \(j\)th order
Hermite polynomial
\citetext{\citealp{hermite1864nouveau}; \citealp[originally
by][]{de1820theorie}} term of the predictor from the simple linear
regression equation. Four different values of \(j = 2, 3, 6, 18\) were
chosen so that distinct shapes of non-linearity were included in the
residual plots. These include ``U'', ``S'', ``M'' and ``triple-U'' shape
as shown in Figure \ref{fig:different-shape-of-herimite}. A greater
value of \(j\) will result in a curve with more turning points. It is
expected that the ``U'' shape will be the easiest to detect, and as the
shape gets more complex it will be harder to perceive in a scatterplot,
particularly when there is noise. Figure \ref{fig:different-sigma} shows
the ``U'' shape for different amounts of noise (\(\sigma\)).

\begin{figure}[!h]

{\centering \includegraphics[width=1\linewidth]{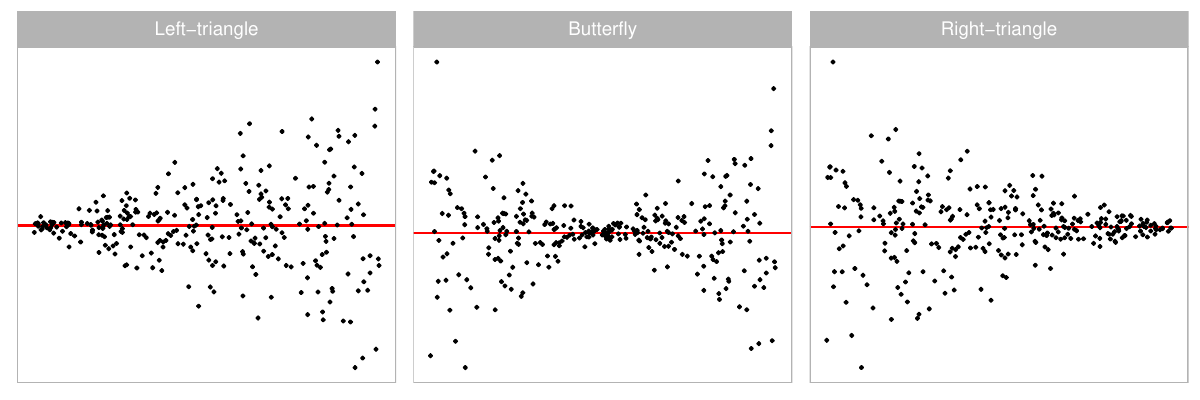} 

}

\caption{Heteroskedasticity forms used in the experiment. Three different shapes ($a = -1, 0, 1$) are used in the experiment to create ``left-triangle", ``butterfly" and ``right-triangle" shapes, respectively.}\label{fig:different-shape-of-heter}
\end{figure}

\begin{figure}[!h]

{\centering \includegraphics[width=1\linewidth]{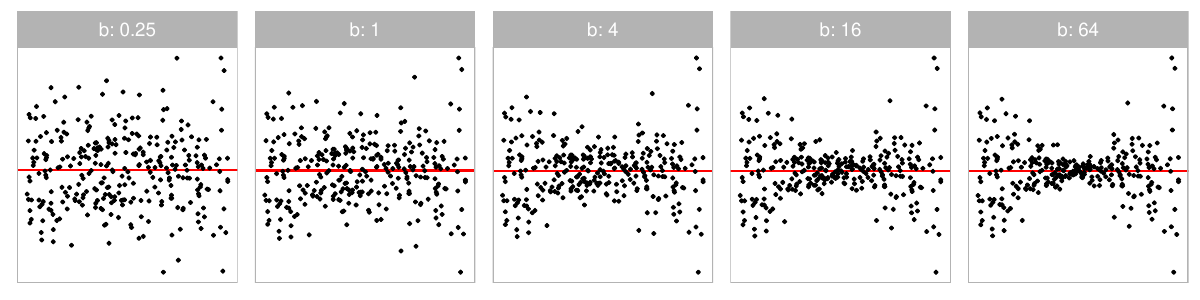} 

}

\caption{Five different values of $b$ are used in heteroskedasticity simulation to control the strength of the signal. Larger values of $b$ yield a bigger difference in variation, and thus stronger heteroskedasticity signal.}\label{fig:different-b}
\end{figure}

Data collection period II was similar to period I but focuses on
heteroskedasticity departures. We generated the heteroskedasticity
departures by setting the variance-covariance matrix of the error term
as a function of the predictor, but fitted the data with the simple
linear regression model, intentionally violated the constant variance
assumption. Visual patterns of heteroskedasticity are simulated using
three different shapes (\(a\) = -1, 0, 1) including ``left-triangle'',
``butterfly'' and ``right-triangle'' shapes as displayed in Figure
\ref{fig:different-shape-of-heter}. Figure \ref{fig:different-b} shows
the butterfly shape as the ratio parameter (\(b\)) is changed. More
details about the simulation process are provided in Appendix B.

\subsubsection{Factors common to both data collection
periods}\label{factors-common-to-both-data-collection-periods}

Fitted values are a function of the independent variables (or
predictors), and the distribution of the observed values affects the
distribution of the fitted values. Ideally, we would want the fitted
values to have a uniform coverage across the range of observed values or
have a uniform distribution across all of the predictors. This is not
always present in the collected data. Sometimes the fitted values are
discrete because one or more predictors were measured discretely. It is
also common to see a skewed distribution of fitted values if one or more
of the predictors has a skewed distribution. This latter problem is
usually corrected before modelling, using a variable transformation. Our
simulation assess this by using four different distributions to
represent fitted values, constructed by different sampling of the
predictor, including \(U(-1, 1)\) (uniform), \(N(0, 0.3^2)\) (normal),
\(\text{lognormal}(0, 0.6^2)/3\) (skewed) and \(U\{-1, 1\}\) (discrete).

\begin{figure}

{\centering \includegraphics[width=1\linewidth]{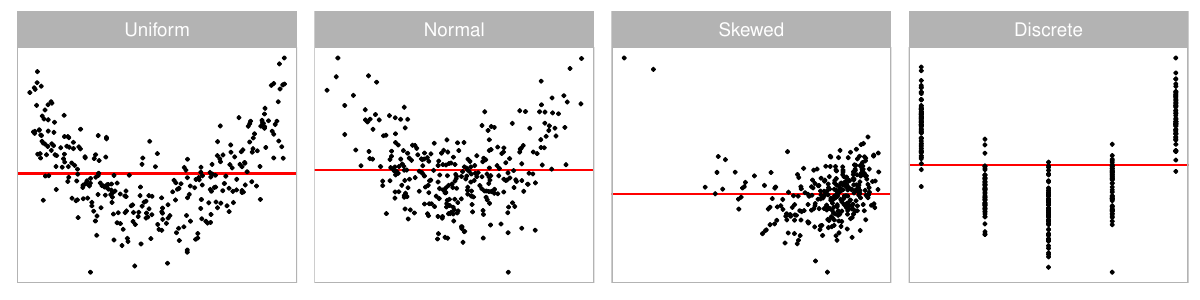} 

}

\caption{Variations in fitted values, that might affect perception of residual plots. Four different distributions are used.}\label{fig:different-dist}
\end{figure}

Figure \ref{fig:different-dist} shows the non-linear pattern, a ``U''
shape, with the different fitted value distributions. We would expect
that structure in residual plots would be easier to perceive when the
fitted values are uniformly distributed.

\begin{figure}

{\centering \includegraphics[width=1\linewidth]{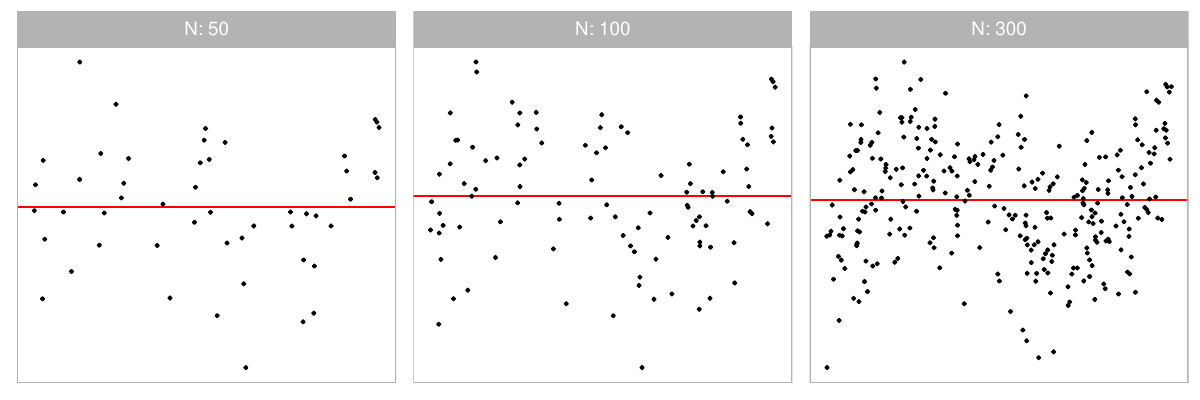} 

}

\caption{Examining the effect of signal strength for the three different values of $n$ used in the experiment, for non-linear structure with fixed $\sigma = 1.5$, uniform fitted value distribution, and "S" shape. For these factor levels, only when $n = 300$ is the "S" shape clearly visible.}\label{fig:different-n}
\end{figure}

Three different sample sizes were used in our experiment:
\(n = 50, 100, 300\). Figure \ref{fig:different-n} shows the non-linear
``S'' shape for different sample sizes. We expect signal strength to
decline in the simulated data plots with smaller \(n\). We chose 300 as
the upper limit, because it is typically enough for structure to be
visible in a scatterplot reliably. Beyond 300, the scatterplot should
probably be used with transparency or replaced with a density or binned
plot as scatterplots suffer from over-plotting.

\subsection{Effect size}\label{effect-size}

The lineups are allocated to participants in a manner that uniformly
covers the combination of experimental factors in Table
\ref{tab:model-factor-table}. In addition, we use effect size to measure
the signal strength, which helps in assigning a set of lineups with a
range of difficulties to each participant.

Effect size in statistics measures the strength of the signal relative
to the noise. It is surprisingly difficult to quantify, even for
simulated data as used in this experiment.

For the non-linearity model, the key items defining effect size are
sample size (\(n\)) and the noise level (\(\sigma^2\)), and so effect
size would be roughly calculated as \(\sqrt{n}/{\sigma}\). Increasing
sample size tends to boost the effect size, while heightened noise
diminishes it. However, it is not clear how the additional parameter for
the model polynomial order, \(j\), should be incorporated. Intuitively,
the large \(j\) means more complex pattern, which likely means effect
size would decrease. Similarly, in the heteroskedasticity model, effect
size relies on sample size (\(n\)) and the ratio of the largest to
smallest variance, \(b\). Larger values of both would produce higher
effect size, but the role of the additional shape parameter, \(a\), in
this context is unclear.

For the purposes of our calculations we have chosen to use an approach
based on Kullback-Leibler divergence \citep{kullback1951information}.
This formulation defines effect size to be

\vspace{-\baselineskip}

\[E = \frac{1}{2}\left(\log\frac{|\text{diag}(\boldsymbol{R}\boldsymbol{V}\boldsymbol{R}')|}{|\text{diag}(\boldsymbol{R}\sigma^2)|} - n + \text{tr}(\text{diag}(\boldsymbol{R}\boldsymbol{V}\boldsymbol{R}')^{-1}\text{diag}(\boldsymbol{R}\sigma^2)) + \boldsymbol{\mu}_z'(\boldsymbol{R}\boldsymbol{V}\boldsymbol{R}')^{-1}\boldsymbol{\mu}_z\right),\]

\noindent where \(\text{diag}(.)\) is the diagonal matrix constructed
from the diagonal elements of a matrix, \(\boldsymbol{X}\) is the design
matrix, \(\boldsymbol{V}\) is the actual covariance matrix of the error
term,
\(\boldsymbol{R} = \boldsymbol{I}_n - \boldsymbol{X}(\boldsymbol{X}'\boldsymbol{X})^{-1}\boldsymbol{X}'\)
is the residual operator,
\(\boldsymbol{\mu}_z = \boldsymbol{R}\boldsymbol{Z}\boldsymbol{\beta}_z\)
is the expected values of residuals where \(\boldsymbol{Z}\) contains
any higher order terms of \(\boldsymbol{X}\) left out of the regression
equation, \(\boldsymbol{\beta}_z\) contains the corresponding
coefficients, and \(\sigma^2\boldsymbol{I}_n\) is the assumed covariance
matrix of the error term when \(H_0\) is true. More details about the
effect size derivation are provided in Appendix A.

\section{Results}\label{results}

Data collection used a total of 1152 lineups, and resulted in a total of
7974 evaluations from 443 participants. Roughly half corresponded to the
two models, non-linearity and heteroskedasticiy, and the three
collection periods had similar numbers of evaluations. Each participant
received two of the 24 attention check lineups which were used to filter
results of participants who were clearly not making an honest effort
(only 11 of 454). To estimate \(\alpha\) for calculating statistical
significance (see Section A.1 in the Appendix) there were 720
evaluations of 36 null lineups. Neither the attention checks nor null
lineups were used in the subsequent analysis. The de-identified data,
\texttt{vi\_survey}, is made available in the R package,
\texttt{visage}.

The data was collected on lineups constructed from four different fitted
value distributions that stem from the corresponding predictor
distribution: uniform, normal, skewed and discrete. Henceforth, we refer
to these four different fitted value distributions with respect to their
predictor distribution. More data was collected on the uniform
distribution (each evaluated by 11 participants) than the others (each
evaluated by 5 participants). The analysis in Sections
\ref{power-analysis}--\ref{hetero-analysis} uses only results from
lineups with uniform distribution, for a total 3069 lineup evaluations.
This allows us to compare the conventional and visual test performance
in an optimal scenario. Section
\ref{effect-of-fitted-value-distributions} examines how the results may
be affected if the fitted value distribution was different.

\subsection{\texorpdfstring{Power comparison of the
tests\label{power-analysis}}{Power comparison of the tests}}\label{power-comparison-of-the-tests}

\begin{figure}

{\centering \includegraphics[width=1\linewidth]{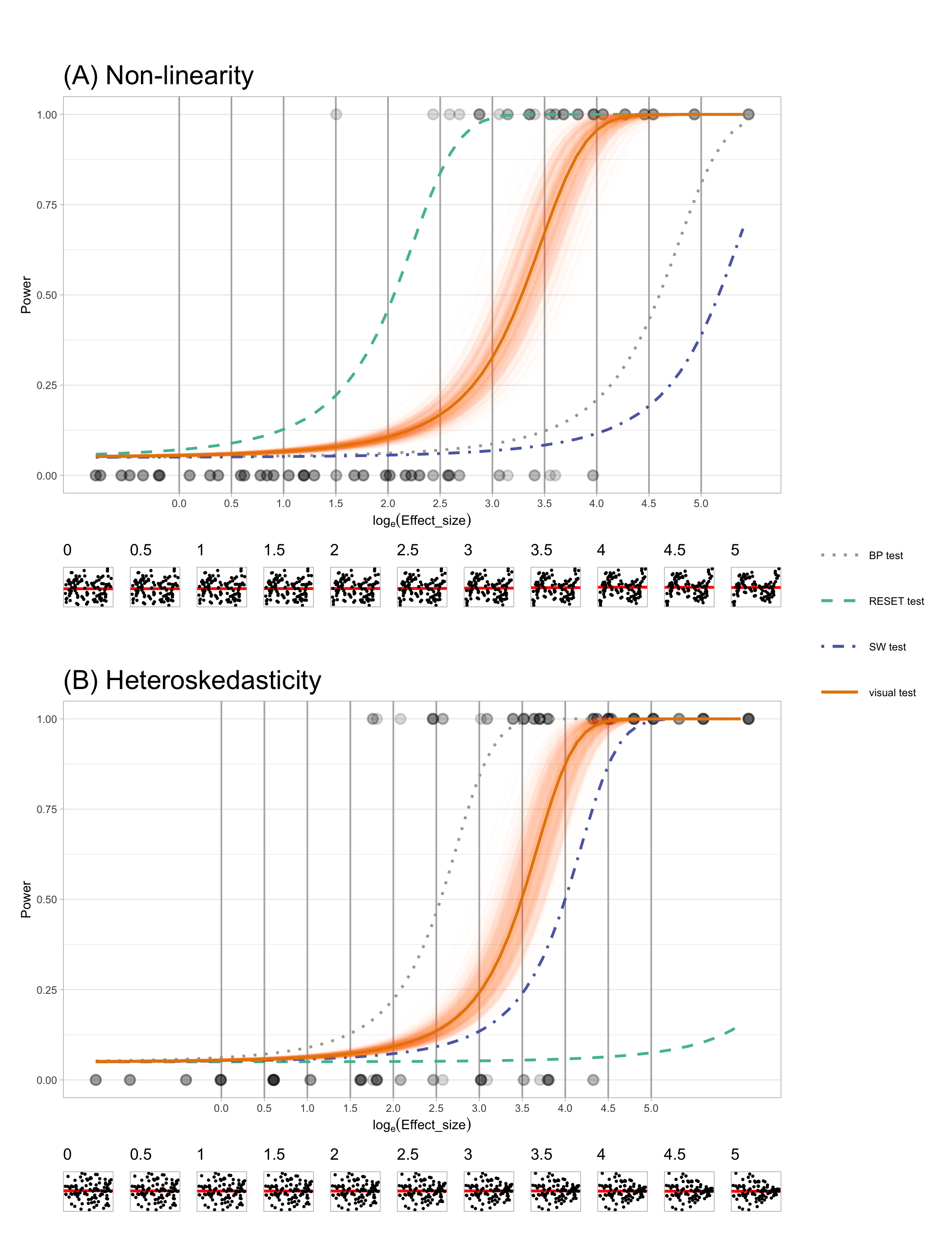} 

}

\caption{Comparison of power between different tests for (A) non-linear and (B) heteroskedasticity patterns (uniform fitted values only). Main plot shows the power curves, with dots indicating non-reject and reject in visual testing of lineups. The multiple lines for the visual test arise from estimating the power on many bootstrap samples. The row of scatterplots at the bottom are examples of residual plots corresponding to the specific effect sizes marked by vertical lines in the main plot.}\label{fig:nonlinearheterpower}
\end{figure}

Figure \ref{fig:nonlinearheterpower} present the power curves of various
tests plotted against the effect size in the residuals for non-linearity
and heteroskedasticity. In each case the power of visual test is
calculated for multiple bootstrap samples leading to the many (solid
orange) curves. The effect size was computed at a 5\% significance level
and plotted on a natural logarithmic scale. To facilitate visual
calibration of effect size values with the corresponding diagnostic
plots, a sequence of example residual plots with increasing effect sizes
is provided at the bottom of these figures. These plots serve as a
visual aid to help readers understand how different effect size values
translate to changes in the diagnostic plots. The horizontal lines of
dots at 0 and 1 represent the non-rejection or rejection decisions made
by visual tests for each lineup.

Figure \ref{fig:nonlinearheterpower}A compares the power for the
different tests for non-linear structure in the residuals. The test with
the uniformly higher power is the RESET test, one that specifically
tests for non-linearity. Note that the BP and SW tests have much lower
power, which is expected because they are not designed to detect
non-linearity. The bootstrapped power curves for the visual test are
effectively a right shift from that of the RESET test. This means that
the RESET test will reject at a lower effect size (less structure) than
the visual test, but otherwise the performance will be similar. In other
words, the RESET test is more sensitive than the visual test. This is
not necessarily a good feature for the purposes of diagnosing model
defects: if we scan the residual plot examples at the bottom, we might
argue that the non-linearity is not sufficiently problematic until an
effect size of around 3 or 3.5. The RESET test would reject closer to an
effect size of 2, but the visual test would reject closer to 3.25, for a
significance level of 0.05. The visual test matches the robustness of
the model to (minor) violations of assumptions much better.

For the heteroskedasticity pattern, the power of BP test, designed for
detecting heteroskedasticity, is uniformly higher than the other tests.
The visual test power curve shifts to the right. This shows a similar
story to the power curves for non-linearity pattern: the conventional
test is more sensitive than the visual test. From the example residual
plots at the bottom we might argue that the heteroskedasticity becomes
noticeably visible around an effect size of 3 or 3.5. However the BP
test would reject at around effect size 2.5. Interestingly, the power
curve for the SW test (for non-normality) is only slightly different to
that of the visual test, suggesting that it performs reasonably well for
detecting heteroskedasticity, too. The power curve for the BP test
suggests it is not useful for detecting heteroskedasticity, as expected.

Overall, the results show that the conventional tests are more sensitive
than the visual test. The conventional tests do have higher power for
the patterns they are designed to detect, but they typically fail to
detect other patterns unless those patterns are particularly strong. The
visual test does not require specifying the pattern ahead of time,
relying purely on whether the observed residual plot is detectably
different from ``good'' residual plots. They will perform equally well
regardless of the type of model defect. This aligns with the advice of
experts on residual analysis, who consider residual plot analysis to be
an indispensable tool for diagnosing model problems. What we gain from
using a visual test for this purpose is the removal of any subjective
arguments about whether a pattern is visible or not. The lineup protocol
provides the calibration for detecting patterns: if the pattern in the
data plot cannot be distinguished from the patterns in good residual
plots, then no discernible problem with the model exists.

\subsection{\texorpdfstring{Comparison of test decisions based on
\(p\)-values\label{p-value}}{Comparison of test decisions based on p-values}}\label{comparison-of-test-decisions-based-on-p-values}

\begin{figure}

{\centering \includegraphics[width=1\linewidth]{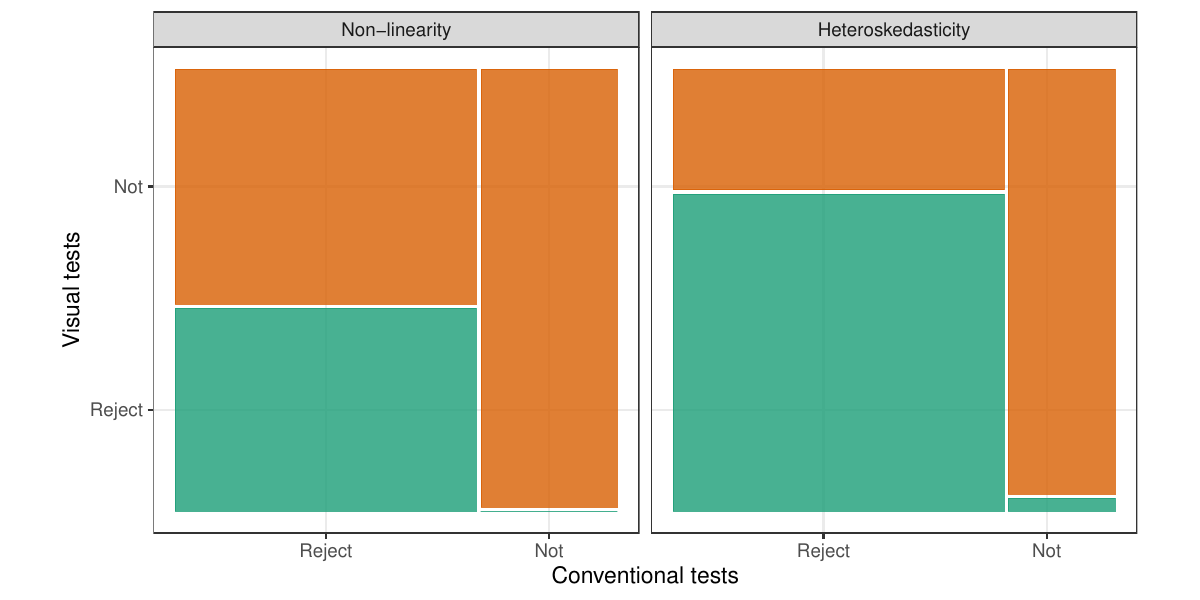} 

}

\caption{Rejection rate ($p$-value $\leq0.05$) of visual test conditional on the conventional test decision on non-linearity (left) and heteroskedasticity (right) lineups (uniform fitted values only) displayed using a mosaic plot. The visual test rejects less frequently than the conventional test, and (almost) only rejects when the conventional test does. Surprisingly, one lineup in the heteroskedasticity group is rejected by the visual test but NOT the conventional test.}\label{fig:p-value-comparison}
\end{figure}

The power comparison demonstrates that the appropriate conventional
tests will reject more aggressively than visual tests, but we do not
know how the decisions for each lineup would agree or disagree. Here we
compare the reject or fail to reject decisions of these tests, across
all the lineups. Figure \ref{fig:p-value-comparison} shows the agreement
of the conventional and visual tests using a mosaic plot for both
non-linearity patterns and heteroskedasticity patterns. For both
patterns the lineups resulting in a rejection by the visual test are
\emph{all} also rejected by the conventional test, except for one from
the heteroskedasticity model. This reflects exactly the story from the
previous section, that the conventional tests reject more aggressively
than the visual test.

For non-linearity lineups, conventional tests and visual tests reject
69\% and 32\% of the time, respectively. Of the lineups rejected by the
conventional test, 46\% are rejected by the visual test, that is,
approximately half as many as the conventional test. There are no
lineups that are rejected by the visual test but not by the conventional
test.

In heteroskedasticity lineups, 76\% are rejected by conventional tests,
while 56\% are rejected by visual tests. Of the lineups rejected by the
conventional test, the visual test rejects more than two-thirds of them,
too.

\begin{figure}[t!]

{\centering \includegraphics[width=1\linewidth]{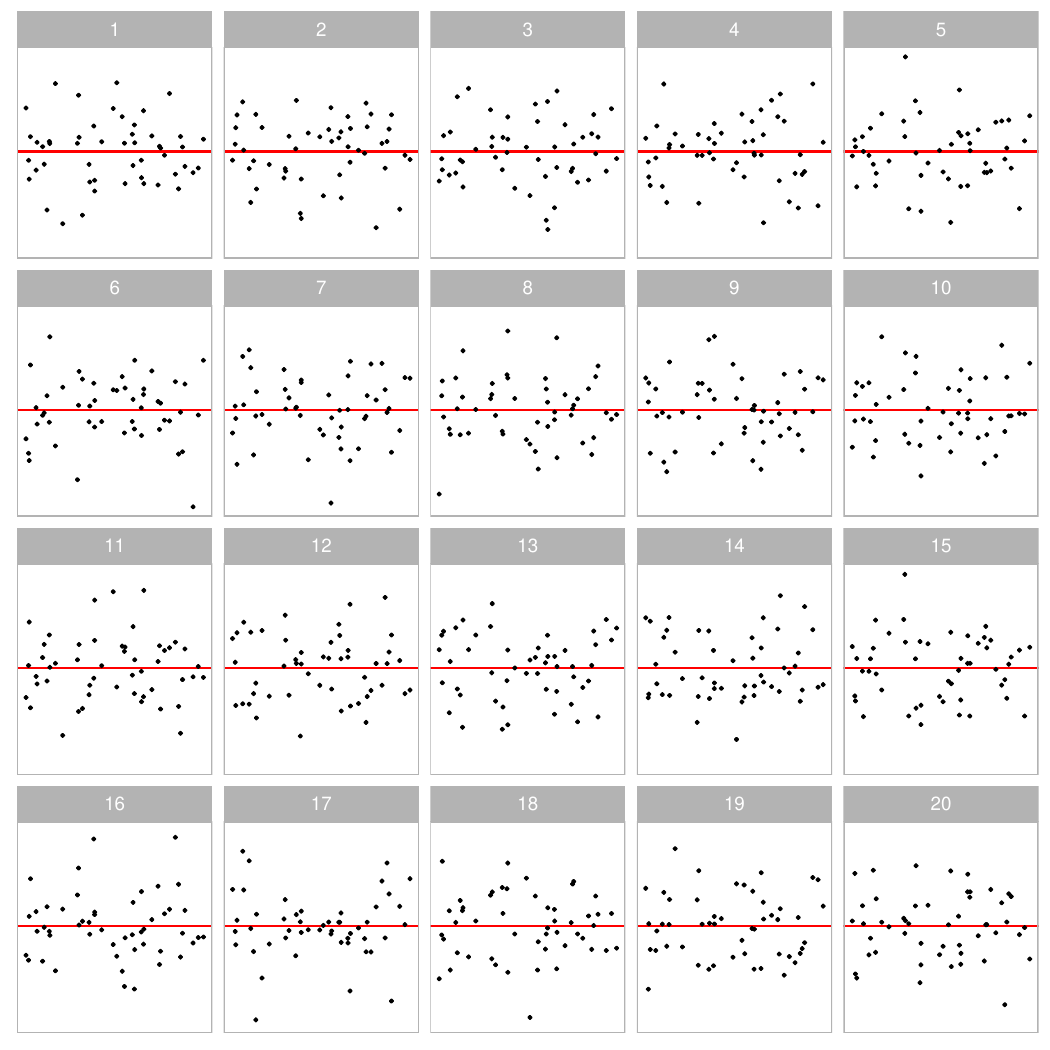} 

}

\caption{The single heteroskedasticity lineup that is rejected by the visual test but not by the BP test. The data plot (position 17) contains a ``butterfly" shape. It visibly displays heteroskedasticity, making it somewhat surprising that it is not detected by the BP test.}\label{fig:heter-example}
\end{figure}

Surprisingly, the visual test rejects 1 of the 33 (3\%) of lineups where
the conventional test does not reject. Figure \ref{fig:heter-example}
shows this lineup. The data plot in position seventeen displays a
relatively strong heteroskedasticity pattern, and has a strong effect
size (\(\log_e(E)=4.02\)), which is reflected by the visual test
\(p\text{-value} = 0.026\). But the BP test \(p\text{-value} = 0.056\),
is slightly above the significance cutoff of \(0.05\). This lineup was
evaluated by 11 participants, it has experimental factors \(a = 0\)
(``butterfly'' shape), \(b = 64\) (large variance ratio), \(n = 50\)
(small sample size), and a uniform distribution for the predictor. It
may have been the small sample size and the presence of a few outliers
that may have resulted in the lack of detection by the conventional
test.

Because the power curve of the visual tests are a shift to the right of
the conventional test (Figure \ref{fig:nonlinearheterpower}) we examined
whether adjusting the significance level (to .001, .0001, .00001,
\ldots) of the conventional test would generate similar decisions to
that of the visual test. Interestingly, it does not: despite resulting
in less rejections, neither the RESET or BP tests come to complete
agreement with the visual test (see Appendix A).

\subsection{\texorpdfstring{Effect of amount of
non-linearity\label{nonlin-analysis}}{Effect of amount of non-linearity}}\label{effect-of-amount-of-non-linearity}

\begin{figure}

{\centering \includegraphics[width=1\linewidth]{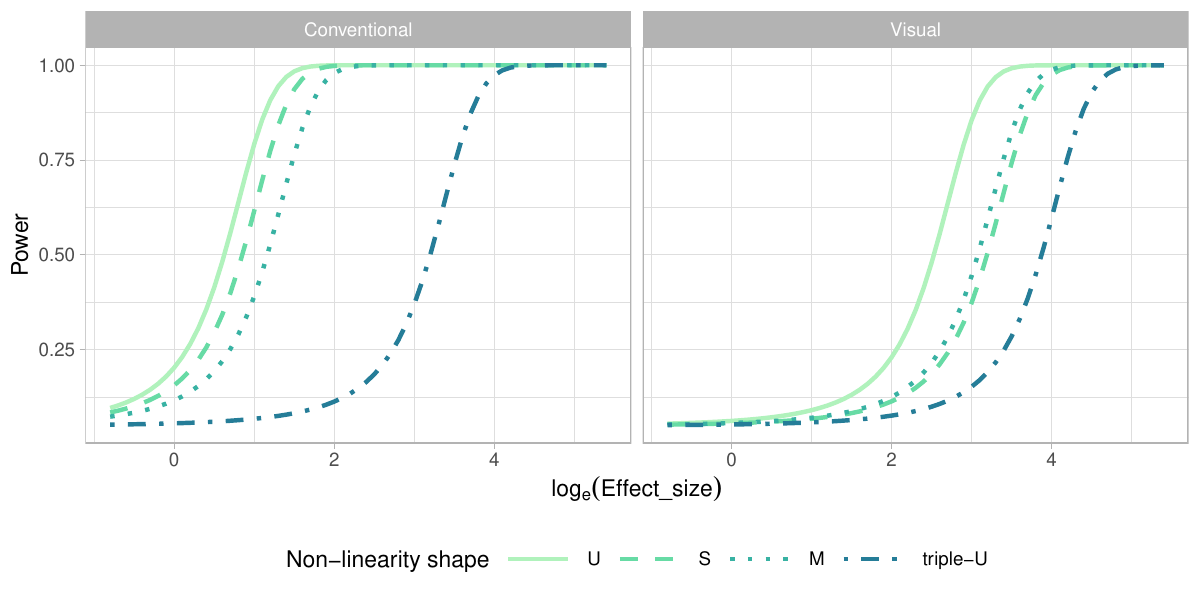} 

}

\caption{The effect of the order of the polynomial on the power of conventional and visual tests. Deeper colour indicates higher order. The default RESET tests under-performs significantly in detecting the "triple-U" shape. To achieve a similar power as other shapes, a higher order polynomial parameter needs to be used for the RESET test, but this higher than the recommended value.}\label{fig:poly-power-uniform-j}
\end{figure}

The order of the polynomial is a primary factor contributing to the
pattern produced by the non-linearity model. Figure
\ref{fig:poly-power-uniform-j} explores the relationship between
polynomial order and power of the tests. The conventional tests have
higher power for lower orders of Hermite polynomials, and the power
drops substantially for the ``triple-U'' shape. To understand why this
is, we return to the application of the RESET test, which requires a
parameter indicating degree of fitted values to test for, and the
recommendation is to generically use four \citep{ramsey1969tests}.
However, the ``triple-U'' shape is constructed from the Hermite
polynomials using power up to 18. If the RESET test had been applied
using a higher power of no less than six, the power curve of
``triple-U'' shape will be closer to other power curves. This
illustrates the sensitivity of the conventional test to the parameter
choice, and highlights a limitation: it helps to know the data
generating process to set the parameters for the test, which is
unrealistic in practice. However, we examined this in more detail (see
Appendix A) and found that there is no harm in setting the parameter
higher than four on the tests' operation for lower order polynomial
shapes. Using a parameter value of six, instead of four, yields higher
power regardless of generating process, and is recommended.

For visual tests, we expect the ``U'' shape to be detected more readily,
followed by the ``S'', ``M'' and ``triple-U'' shape. From Figure
\ref{fig:poly-power-uniform-j}, it can be observed that the power curves
mostly align with these expectations, except for the ``M'' shape, which
is as easily detected as the ``S'' shape. This suggests a benefit of the
visual test: knowing the shape ahead of time is \emph{not} needed for
its application.

\subsection{\texorpdfstring{Effect of shape of
heteroskedasticity\label{hetero-analysis}}{Effect of shape of heteroskedasticity}}\label{effect-of-shape-of-heteroskedasticity}

\begin{figure}

{\centering \includegraphics[width=1\linewidth]{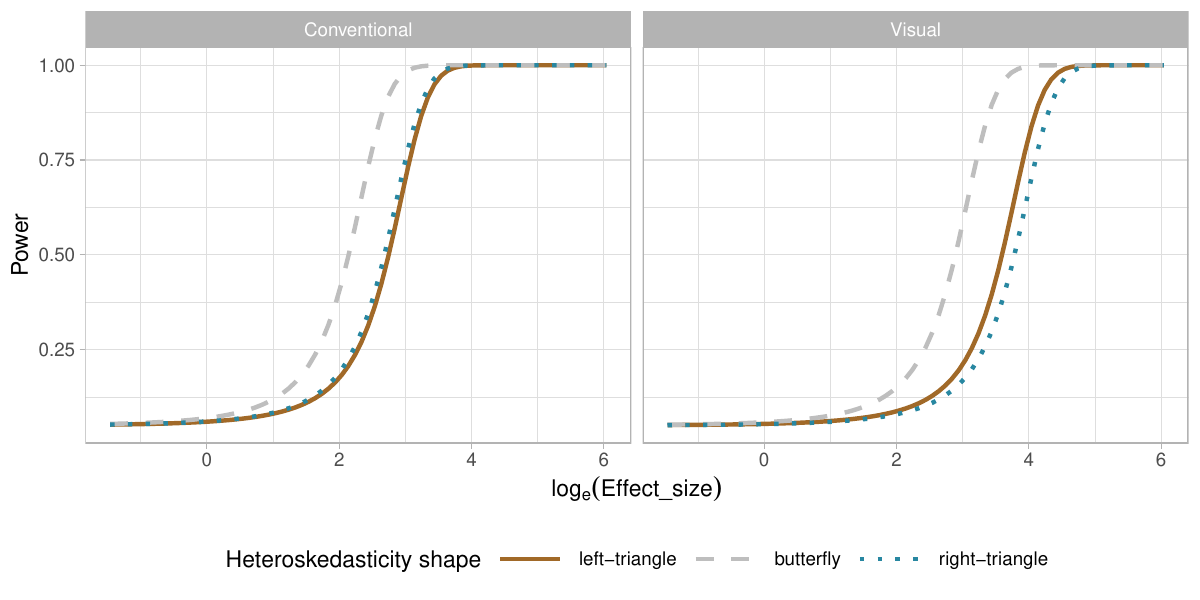} 

}

\caption{The effect of heteroskedasticity shape (parameter $a$) on the power of conventional and visual tests. The butterfly has higher power in both tests. Curiously, the visual test has a slightly higher power for the ``left-triangle" than the ``right-triangle" shape, when it would be expected that they should be similar, which is observed in conventional testing.}\label{fig:heter-power-uniform-a}
\end{figure}

Figure \ref{fig:heter-power-uniform-a} examines the impact of the shape
of the heteroskedasticity on the power of of both tests. The butterfly
shape has higher power on both types of tests. The ``left-triangle'' and
the ``right-triangle'' shapes are functionally identical, and this is
observed for the conventional test, where the power curves are
identical. Interestingly there is a difference for the visual test: the
power curve of the ``left-triangle'' shape is slightly higher than that
of the ``right-triangle'' shape. This indicates a bias in perceiving
heteroskedasticity depending on the direction, and may be worth
investigating further.

\subsection{Effect of fitted value
distributions}\label{effect-of-fitted-value-distributions}

In regression analysis, predictions are conditional on the observed
values of the predictors, that is, the conditional mean of the dependent
variable \(Y\) given the value of the independent variable \(X\),
\(\text{E}(Y|X)\). This is an often forgotten element of regression
analysis but it is important. Where \(X\) is observed, the distribution
of the \(X\) values in the sample, or consequently \(\hat{Y}\), may
affect the ability to read any patterns in the residual plots. The
effect of fitted value distribution on test performance is assess using
four different distributions of fitted values stemming from the
predictor distributions: uniform, normal, discrete and lognormal
(skewed). We expect that if all predictors have a uniform distribution,
it is easier to read the relationship with the residuals.

\begin{figure}[t!]

{\centering \includegraphics[width=1\linewidth]{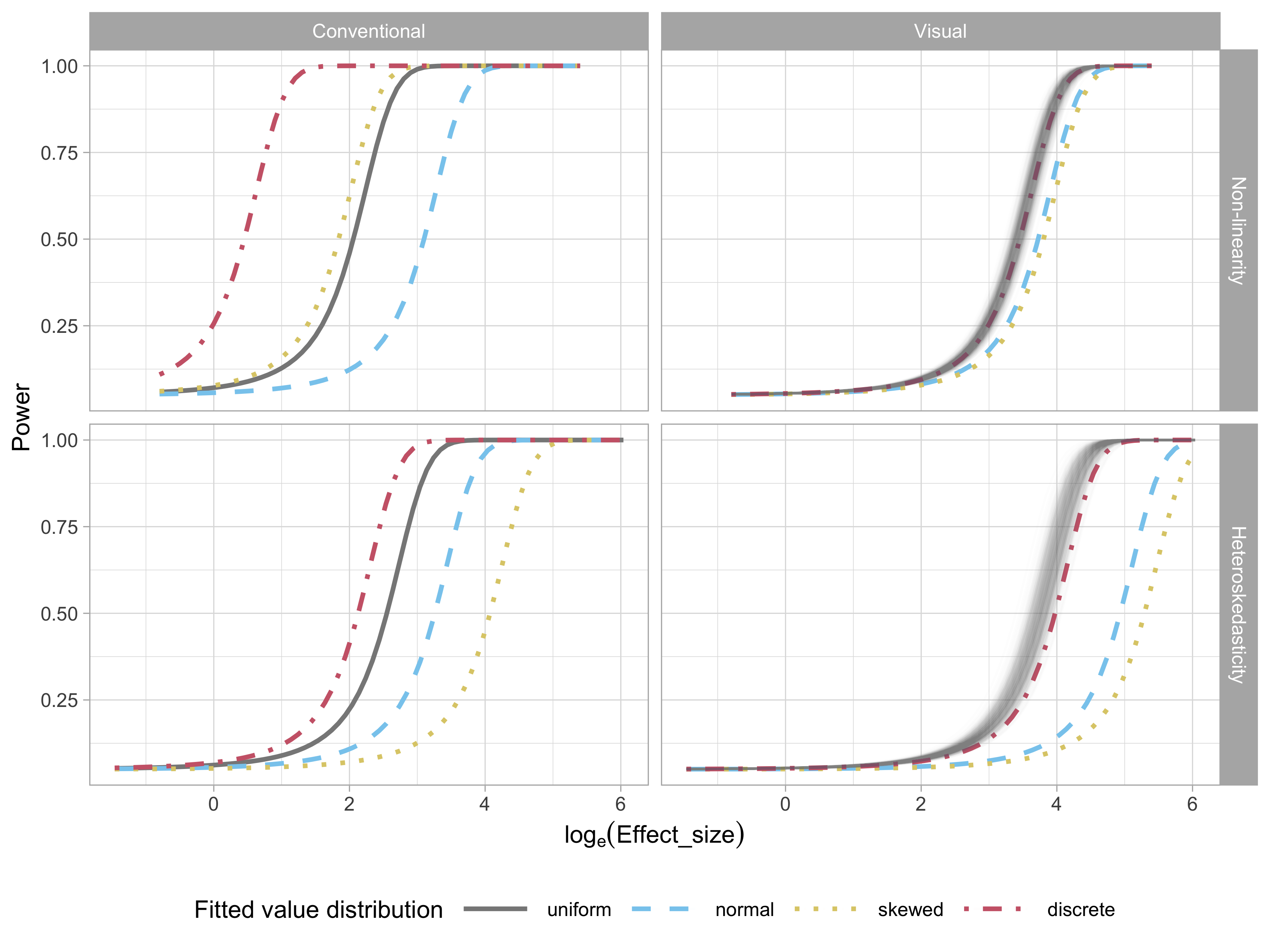} 

}

\caption{Comparison of power on lineups with different fitted value distributions for conventional and visual tests (columns) for non-linearity and heteroskedasticity patterns (rows). The power curves of conventional tests for non-linearity and heteroskedasticity patterns are produced by RESET tests and BP tests, respectively. Power curves of visual tests are estimated using five evaluations on each lineup. For lineups with a uniform fitted value distribution, the five evaluations are repeatedly sampled from the total eleven evaluations to give multiple power curves (solid grey). Surprisingly, the fitted value distribution has produces more variability in the power of conventional tests than visual tests. Uneven distributions, normal and skewed distributions, tend to yield lower power.}\label{fig:different-x-dist-poly-power}
\end{figure}

Figure \ref{fig:different-x-dist-poly-power} examines the impact of the
fitted value distribution on the power of conventional (left) and visual
(right) tests for both the non-linearity (top) and heteroskedasticity
(bottom) patterns. For conventional tests, only the power curves of
appropriate tests are shown: RESET tests for non-linearity and BP tests
for heteroskedasticity. For visual tests, more evaluations on lineups
with uniform fitted value distribution were collected, so to have a fair
comparison, we randomly sample five from the 11 total evaluations to
estimate the power curves, producing the multiple curves for the uniform
condition, and providing an indication of the variability in the power
estimates.

Perhaps surprisingly, the visual tests have more consistent power across
the different fitted value distributions: for the non-linear pattern,
there is almost no power difference, and for the heteroskedastic
pattern, uniform and discrete have higher power than normal and skewed.
The likely reason is that these latter two have fewer observations in
the tails where the heteroskedastic pattern needs to be detected.

The variation in power in the conventional tests is at first sight,
shocking. However, it is discussed, albeit rarely, in the testing
literature. See, for example, \citet{jamshidian2007study},
\citet{olvera2019relationship} and \citet{zhang2018practical} which show
derivations and use simulation to assess the effect of the observed
distribution of the predictors on test power. The big differences in the
power curves seen in Figure \ref{fig:different-x-dist-poly-power} is
echoed in the results reported in these articles.

\section{Limitations and
practicality}\label{limitations-and-practicality}

One of the primary limitations of the lineup protocol lies in its
reliance on human judgements. In this context, the effectiveness of a
single lineup evaluation can be dependent on the perceptual ability and
visual skills of the individual. However, when results from multiple
individuals are combined the outcome is encouragingly high-quality and
robust. For simple plots and strong patterns just a few individuals are
needed to arrive at a clear answer, but more individuals will be needed
when the plot design is complex, or the signal strength is weak.

Using a lineup protocol removes subjectiveness in interpreting patterns
in plots. A plot is compared with draws from a null model, in much the
same way as a test statistic is compared to its sampling distribution.
It is important to remove plot elements that might introduce bias, such
as axis labels, text and legends, or to make them generic.

The lineup protocol can be used cheaply and informally with the R
package \texttt{nullabor}. There is evidence that it is being used
fairly broadly, based on software download rates and citations of the
original papers. For residual plot analysis we recommend that the lineup
be the default first plot so that the data plot is only seen in the
context of null plots. When a rigorous test is needed, we recommend
using a crowd-sourcing service, as done in gene expression experiment
described in \citet{RNAseq2013}. While it takes extra effort it is not
difficult today, and costs are tiny compared to the overall costs of
conducting a scientific experiment. We do also expect that at some point
a computer vision model can be developed to take over the task of
employing people to evaluate residual plots.

For this study, simulated data was used to provide a precisely
controlled environment within which to compare results from conventional
testing to those from visual testing. We also explored only the most
commonly used, the residual vs fitted value plots. However, we expect
the behavior of the conventional test and the visual test to be similar
when observed residuals are diagnosed with this type of plot or other
residual plots. The conventional tests will be more sensitive to small
departures from the null. They will also fail to detect departures when
residuals have some contamination, like outliers or anomalies, as is
often encountered when working with data. The lineup approach is
well-suited for generally interpreting data plots, and also detecting
unexpected patterns not related to the model. This is supported by
earlier research
\citep[e.g.][]{wickham2010, roy2015using, loy2015you, VanderPlas2015, loy2016variations}.

\section{Conclusions}\label{conclusions}

This paper has described experimental evidence providing support for the
advice of regression analysis experts \emph{that residual plots are
indispensable methods for assessing model fit}, using the formal
framework of the lineup protocol. We conducted a perceptual experiment
on scatterplots of residuals vs fitted values, with two primary
departures from good residuals: non-linearity and heteroskedasticity. We
found that conventional residual-based statistical tests are more
sensitive to weak departures from model assumptions than visual tests.
That is, a conventional test concludes there are problems with the model
fit almost twice as often as a human. Conventional tests often reject
the null hypothesis when departures in the form of non-linearity and
heteroskedasticity are not visibly different from null residual plots.

While it might be argued that the conventional tests are correctly
detecting small but real effects, this can also be seen as the
conventional tests are rejecting unnecessarily. Many of these rejections
happen even when downstream analysis and results would not be
significantly affected by the small departures from a good fit. The
results from human evaluations provide a more practical solution, which
reinforces the statements from regression experts that residual plots
are an indispensable method for model diagnostics. Further work would be
needed to \emph{quantify} how much departure from good residuals is too
much.

It is important to emphasize that this work also supports a change in
common practice, which is to deliver residual plots as a lineup,
embedded in a field of null plots, rather than be viewed out of context.
A residual plot may contain many visual features, but some are caused by
the characteristics of the predictors and the randomness of the error,
not by the violation of the model assumptions. These irrelevant visual
features have a chance to be filtered out by participants with a
comparison to null plots, resulting in more accurate reading. The lineup
enables a careful calibration for reading structure in residual plots,
and also provides the potential to discover interesting and important
features in the data not directly connected to linear model assumptions.

Human evaluation of residuals is expensive, time-consuming and
laborious. This is possibly why residual plot analysis is often not done
in practice. However, with the emergence of effective computer vision,
it is hoped this work helps to lay the foundation for automated residual
plot assessment.

The experiment also revealed some interesting results. For the most
part, the visual test performed similarly to the appropriate
conventional test with a shift in the power curve. Unlike conventional
tests, where one needs to specifically test for non-linearity or
heteroskedasticity the visual test operated effectively across the range
of departures from good residuals. If the fitted value distribution is
not uniform, there is a small loss of power in the visual test.
Surprisingly, there is a big difference in power of the conventional
test across fitted value distributions. Another unexpected finding was
that the direction of heteroskedasticity appears to affect the ability
to visually detect it: both triangles being more difficult to detect
than the butterfly, and a small difference in detection between left-
and right-triangle.

\section*{Acknowledgements}\label{acknowledgements}
\addcontentsline{toc}{section}{Acknowledgements}

These \texttt{R} packages were used for the work: \texttt{tidyverse}
\citep{tidyverse}, \texttt{lmtest} \citep{lmtest}, \texttt{mpoly}
\citep{mpoly}, \texttt{ggmosaic} \citep{ggmosaic}, \texttt{kableExtra}
\citep{kableextra}, \texttt{patchwork} \citep{patchwork},
\texttt{rcartocolor} \citep{rcartocolor}. The study website was powered
by \texttt{PythonAnywhere} \citep{pythonanywhere} and \texttt{Flask} web
framework \citep{flask}. The \texttt{jsPsych} framework \citep{jspsych}
was used to create behavioural experiments that run in our study
website.

The article was created with R packages \texttt{rticles}
\citep{rticles}, \texttt{knitr} \citep{knitr} and \texttt{rmarkdown}
\citep{rmarkdown}. The project's GitHub repository
(\url{https://github.com/TengMCing/lineup_residual_diagnostics})
contains all materials required to reproduce this article.

\section*{Supplementary materials}\label{supplementary-materials}
\addcontentsline{toc}{section}{Supplementary materials}

\begin{description}
\item{Appendix:} The appendix includes more details about the experiment setup, the derivation of the effect size, the effect of data collection period, and the estimate of $\alpha$. (appendix.pdf, Portable Document Format file)
\item{R package ``visage":} The R package ``visage" containing code to simulate lineups used in the experiment. The package also contains de-identified data collected from the experiment. (visage\_0.1.2.tar.gz, GNU zipped tar file)
\end{description}

\bibliographystyle{tfcad}
\bibliography{ref.bib}

\end{document}